\documentclass[12pt]{article}
\pdfoutput=1
\usepackage{hyperref}
\usepackage{graphicx}
\usepackage{subcaption}
\graphicspath{{./picture/}}
\usepackage{empheq}
\usepackage{amssymb}
\usepackage{mathtools} 
\usepackage{commath}
\usepackage{verbatim}
\usepackage{xcolor}
\usepackage{soul}
\usepackage{booktabs}
\usepackage{amsmath}
\usepackage{hhline}
\usepackage[scale={0.75,0.8}]{geometry}
\usepackage{url}
\usepackage{caption}
\usepackage[numbers, sort&compress]{natbib}
\usepackage{epsfig}
\usepackage{lscape}
\usepackage{mathrsfs}
\usepackage{nicefrac}
\usepackage{upgreek}
\usepackage{bbm}
\usepackage{psfrag}
\usepackage{hyperref}
\usepackage{cleveref}
\usepackage{hyperref}

\usepackage{enumitem}

\newcommand{\Rll}{R_{ll}}
\newcommand{\pN}{{\rm p}_N}
\newcommand{\cprod}{c_{\rm prod}}
\newcommand{\cdec}{c_{\rm dec}}
\newcommand{\BR}{{\rm B}^{Z}_\alpha}
\renewcommand{\aa}{a}
\newcommand{\NHNL}{N_{{\rm HNL} \alpha}}

\newcommand{\ii}{{\rm i}}
\newcommand{\effa}{\epsilon_{\alpha\beta}}
\newcommand{\ratio}[1]{\mathsf{u}^2_{#1}}
\newcommand{\Nobs}{N_{\rm obs}}
\newcommand{\X}{{\rm X}}
\newcommand{\BRb}{{\rm B}^{N}_{\X_\beta}}

\begin{document}
\title{Distinguishing Dirac and Majorana Heavy Neutrinos\\ at Lepton Colliders
{\bf \boldmath }}
\date{}
\author{Marco Drewes\\ \\ 
	{\normalsize \it $^1$Centre for Cosmology, Particle Physics and Phenomenology,}\\ {\normalsize \it Universit\'{e} catholique de Louvain, Louvain-la-Neuve B-1348, Belgium}\\
}
\maketitle
\thispagestyle{empty}
\begin{abstract}
	\noindent
We discuss the potential to observe lepton number violation (LNV) in displaced vertex searches for heavy neutral leptons (HNLs) at future lepton colliders. Even though a direct detection of LNV is impossible for the dominant production channel because lepton number is carried away by an unobservable neutrino, there are several signatures of LNV that can be searched for. They include the angular distribution and spectrum of decay products as well as the HNL lifetime. We comment on the perspectives to observe LNV in realistic neutrino mass models and argue that the dichotomy of Dirac vs Majorana HNLs is in general not sufficient to effectively capture their phenomenology, but these extreme cases nevertheless represent well-defined benchmarks for experimental searches. Finally, we present accurate analytic estimates for the number of events and sensitivity regions during the $Z$-pole run for both Majorana and Dirac HNLs.  
\end{abstract}


	\tableofcontents
\newpage
\section{Motivation}
Neutrinos are the sole fermions in the Standard Model of particle physics (SM) that could be their own antiparticles, in which case the would be the only known elementary Majorana fermions, 
and their masses would break the global $U(1)_{B-L}$ symmetry of the SM.
An immediate consequence would be the existence of processes that violate the total lepton number $L$. However, due to the smallness of the light neutrino masses $m_i$ the rate for lepton number violating (LNV) processes in neutrino experiments would be parametrically suppressed.\footnote{Neutrinoless double $\beta$-decay can provide an indirect probe \cite{Schechter:1981bd}, cf.~also \cite{Duerr:2011zd,Rodejohann:2011mu}.}
At the same time it is clear that any explanation of the light neutrino masses requires an extension of the SM field content, 
and LNV may occur at an observable rate in processes involving new particles. 
This in particular can include heavy neutral leptons (HNLs)\footnote{
Here we use the following nomenclature:
\emph{HNLs} are fermions with mass $M\gg m_i$ that carry no charge under both the electromagnetic and strong interactions.
\emph{Heavy neutrinos} are a type of HNL that mix with the SM neutrinos.
\emph{Right-handed neutrinos} $\nu_R$ are fields with right-handed chirality that couple to the left-handed SM neutrinos with Yukawa couplings and are singlet (\emph{sterile}) with respect to the SM gauge groups. They are in general not identical to the 
mass eigenstates $N$, cf.~footnote \ref{MassNote}.
In addition to possible connections to neutrino masses, $\nu_R$ can potentially play an important role in other areas of particle physics and cosmology \cite{Abdullahi:2022jlv,Drewes:2013gca}, such as leptogenesis \cite{Fukugita:1986hr}
as an explanation for the observed matter-antimatter asymmetry of the observable universe \cite{Canetti:2012zc} (including low scale scenarios \cite{Akhmedov:1998qx,Pilaftsis:2003gt,Asaka:2005pn,Klaric:2021cpi} that can be tested \cite{Barrow:2022gsu,Chun:2017spz}), or as Dark Matter candidates \cite{Dodelson:1993je,Shi:1998km}.} 
$N_i$ that couple to the $Z$- and $W$-bosons and the Higgs bosons $h$
via the SM weak interaction with an amplitude suppressed by the mixing angles $\theta_{\alpha i}$ (with $\alpha = e,\mu,\tau$ and $i = 1\ldots n $),\footnote{
In general the $N_i$ may have new gauge interactions in addition to the SM weak interactions in \eqref{eq:weak WW} that can also lead to LNV processes (e.g.~\cite{Keung:1983uu}), but the LHC bounds on the mass of new gauge bosons \cite{CMS:2021dzb,CMS:2022irq} make it difficult to explore this option at lepton colliders.
}
\begin{equation}\label{eq:weak WW}
 \mathcal L
\supset
- \frac{m_W}{ v} \overline N_i \theta^*_{\alpha i} \gamma^\mu e_{L \alpha} W^+_\mu
- \frac{m_Z}{\sqrt 2 v} \overline N_i \theta^*_{\alpha i} \gamma^\mu \nu_{L \alpha} Z_\mu
- \frac{M_i}{v\sqrt{2}}  \theta_{\alpha i} h \overline{\nu_L}_\alpha N_i
+ \text{h.c.},
\end{equation}
with $m_Z$, $m_W$ the weak gauge boson masses and $v\simeq 174$ GeV the Higgs field vacuum expectation value.
The $N_i$ can be Dirac or Majorana fermions. 
For $M_i < m_Z$ they can be produced copiously during the $Z$-pole run of future lepton colliders \cite{Blondel:2014bra} such as FCC-ee \cite{FCC:2018evy} or CEPC \cite{CEPCStudyGroup:2018ghi},\footnote{
Linear colliders typically have less sensitivity for $M<m_Z$ \cite{Antusch:2016ejd} due to their smaller integrated luminosity compared to the proposed $Z$-pole runs at FCC-ee or CEPC
\cite{FCC:2018evy,CEPCtalk}, but their polarised beams may offer an advantage when studying forward-backward asymmetries \cite{Hernandez:2018cgc}, cf.~method 1) below.
} 
cf.~Fig.~\ref{1a}, 
making it possible to not only discover them but also study their properties in sufficient detail to probe their role in neutrino mass generation and leptogenesis \cite{Antusch:2017pkq}.
An important question in this context is whether the LNV in $N_i$-decays can be observed. 
This is hampered by two main obstacles, both of which can be overcome,
\begin{itemize}
\item[I)] LNV can be detected most directly when the final state of a process can be fully reconstructed, such as 
    $W^\pm\to\ell_\alpha^\pm  N  \to 
    \ell_\alpha^\pm
    \ell^\pm_\alpha W^\mp_*$. However, at lepton colliders $N_i$ with $M_i < m_Z$ are dominantly produced in the decays of $Z$-bosons along with an unobservable neutrino or antineutrino, making it impossible to reconstruct the final state and determine its total $L$.
    \item[II)] In models that employ the type-I seesaw mechanism \cite{Minkowski:1977sc,GellMann:1980vs, Mohapatra:1979ia,Yanagida:1980xy,Schechter:1980gr,Schechter:1981cv},
    the light neutrino masses parametrically scale as\footnote{\label{MassNote}The type-I seesaw requires the addition of at least $n$ flavours of right-handed neutrinos $\nu_{R}$ with a Majorana mass matrix $M_M$ to the SM in order to generate $n$ non-zero light neutrino masses $m_i$.
The mass eigenstates are represented by Majorana spinors 
$\upnu_i \simeq [U_{\nu}^{\dagger} (\nu_{L} - \theta \nu_{R}^c)]_i + \text{c.c.}$
and
$N_i \simeq [U_N^\dagger ( \nu_{R} +  \theta^T\nu_{L}^{c})] + \text{c.c.}$ with masses $m_i$ and $M_i$, respectively.
The $m_i^2$ and $M_i^2$ at tree level are given by the eigenvalues of $m_\nu m_\nu^\dagger$ and $M_N M_N^\dagger$, with
$M_N  = M_M + \frac{1}{2} (\theta^\dagger \theta M_M + M_M^T \theta^T \theta^{*})$
and
$m_\nu = - \theta M_M \theta^T$. 
$U_\nu$ and $U_N$ diagonalise $m_\nu m_\nu^\dagger$ and $M_N M_N^\dagger$, respectively.
Strictly speaking $\theta$ in \eqref{eq:weak WW} should be replaced by $\Theta = \theta U_N^*$, we neglect this difference for notational simplicity.} $m_i \sim \theta^2 M_i$,
    while the HNL production cross section scales as $\sigma_N \sim \theta^2$, cf.~\eqref{eq:observed events}, so that one may expect $\sigma_N$ to be parametrically suppressed by $\sim m_i/M_i$.\footnote{The precise value of this so-called \emph{seesaw line} in the mass-mixing plane depends on $n$ and the lightest $m_i$ \cite{Drewes:2019mhg}. If all eigenvalues of $M_M$ have a similar magnitude $M$, one can roughly estimate the minimal mixing to be $\simeq \zeta \Delta m_{\rm atm}/M_i$, with $\zeta=1(2)$ for normal (inverted) ordering of the $m_i$ and $\Delta m^2_{\rm atm} \simeq 2.5\times10^{-3} {\rm eV}^2$.} 
    This is not the case if the $m_i$ are protected by an approximate global $U(1)_{B-\bar{L}}$ symmetry, with $\bar{L}$ a generalised lepton-number under which the HNLs are charged \cite{Kersten:2007vk}.  
    The symmetry would lead to systematic cancellations in the neutrino mass matrix that keep the $m_i$ small while allowing for (almost) arbitrarily large
    \begin{eqnarray}\nonumber
    U_{\alpha i}^2 = |\theta_{\alpha i}|^2.
    \end{eqnarray}
    The approximate $\bar{L}$-conservation would, however, also suppress all LNV processes parametrically.  One may expect that the ratio of $L$-violating to $L$-conserving $N_i$-decays scales as 
    $\Rll\sim U_i^{-2} m_i/M_i$
with $U_i^2=\sum_\alpha U_{\alpha i}^2$
and is practically unobservable even if the $N_i$ are fundamentally Majorana particles.
\end{itemize}

\section{Observables sensitive to LNV}\label{sec:Observables}
\begin{figure}
\centering
\vspace{-2cm}
\begin{subfigure}{\textwidth}
\centering
\includegraphics[scale=1.4]{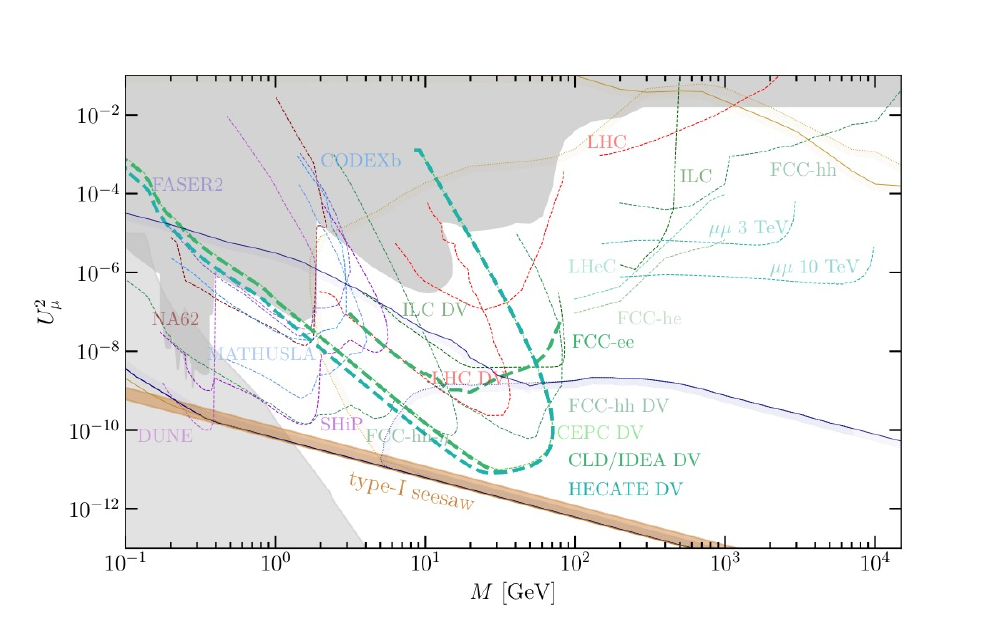}
\caption{\label{1a}}
\end{subfigure}\\
\begin{subfigure}{\textwidth}
\centering
\includegraphics[scale=1]{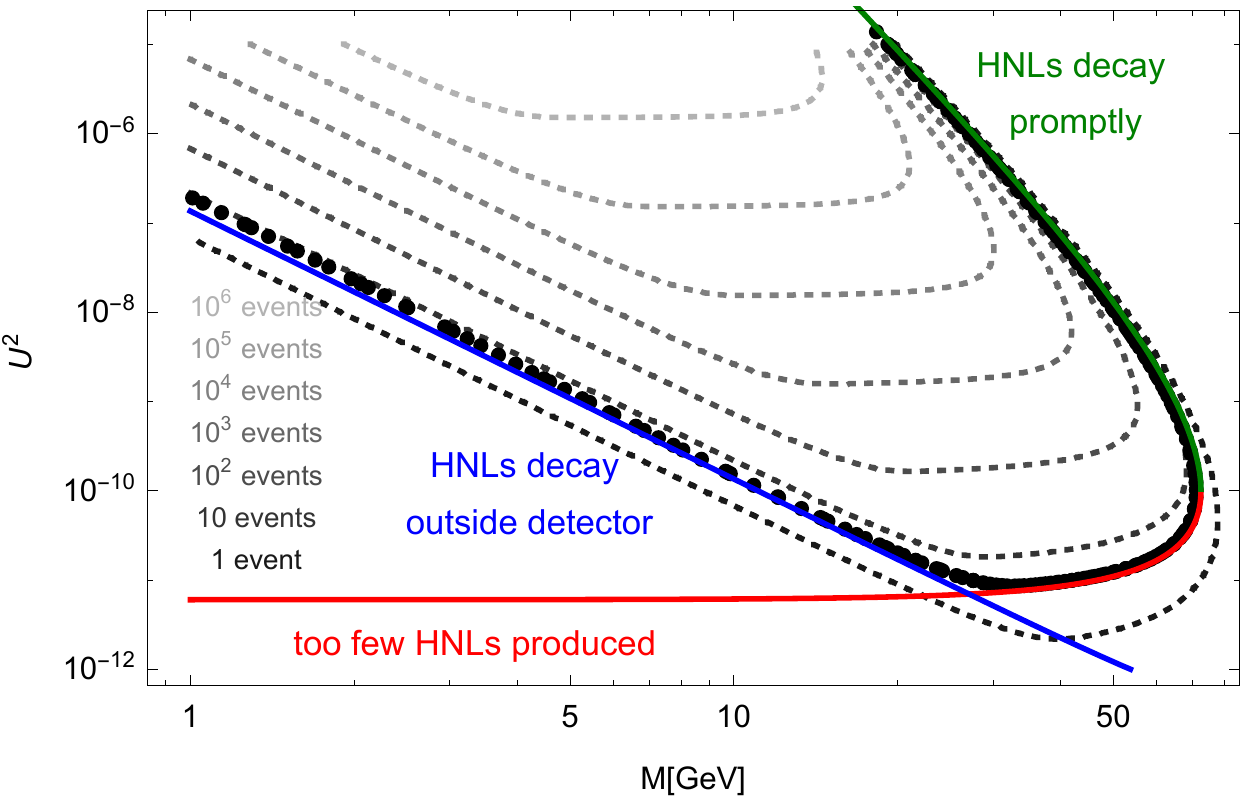}
\caption{\label{1b}}
\end{subfigure}
\caption[\emph{(a)}: Allowed HNL parameter region (white) compared to the sensitivities of various experiments, in particular displaced vertex searches at FCC-ee and CEPC, details given in \cite{Blondel:2022qqo}.
\emph{(b)}: Simulated 4-event curves from \cite{Blondel:2022qqo} (black dots) compared to the analytic estimates \eqref{eq:observed events} (gray dotted), \eqref{RedLine} (red) \eqref{GreenLine} (green) 
and \eqref{BlueLine} (blue)
for 
$U^2=U_\mu^2$, $\cprod=\cdec=1$, $N_\text{obs} = 4$, $N_{\rm IP}=2$, $N_Z = 2.5 \times 10^{12}$, $l_0 = 400\mu$m, $d_{\rm cyl}=10$m, $l_{\rm cyl}=8.6$m, $\effa=1$. 
]{(a) Allowed HNL parameter region (white) compared to the sensitivities of various experiments, in particular displaced vertex searches at FCC-ee and CEPC, details given in \cite{Blondel:2022qqo}.
(b) Simulated 4-event curves from \cite{Blondel:2022qqo} (black dots) compared to the analytic estimates \eqref{eq:observed events} (gray dotted), \eqref{RedLine} (red) \eqref{GreenLine} (green) 
and \eqref{BlueLine} (blue)
for 
$U^2=U_\mu^2$, $\cprod=\cdec=1$, $N_\text{obs} = 4$, $N_{\rm IP}=2$, $N_Z = 2.5 \times 10^{12}$, $l_0 = 400\mu$m, $d_{\rm cyl}=10$m, $l_{\rm cyl}=8.6$m, $\effa=1$.\footnotemark 
Public codes to estimate the number of events and sensitivity regions based on the analytic approximations 
\eqref{eq:observed events}, \eqref{RedLine}, \eqref{GreenLine} 
and \eqref{BlueLine}
are made available at \cite{GitHub}.
} \label{fig:illustrative plot}
\end{figure}

\begin{figure}[h]
\begin{center}
\begin{subfigure}{0.5\textwidth}
\includegraphics[scale=0.6]{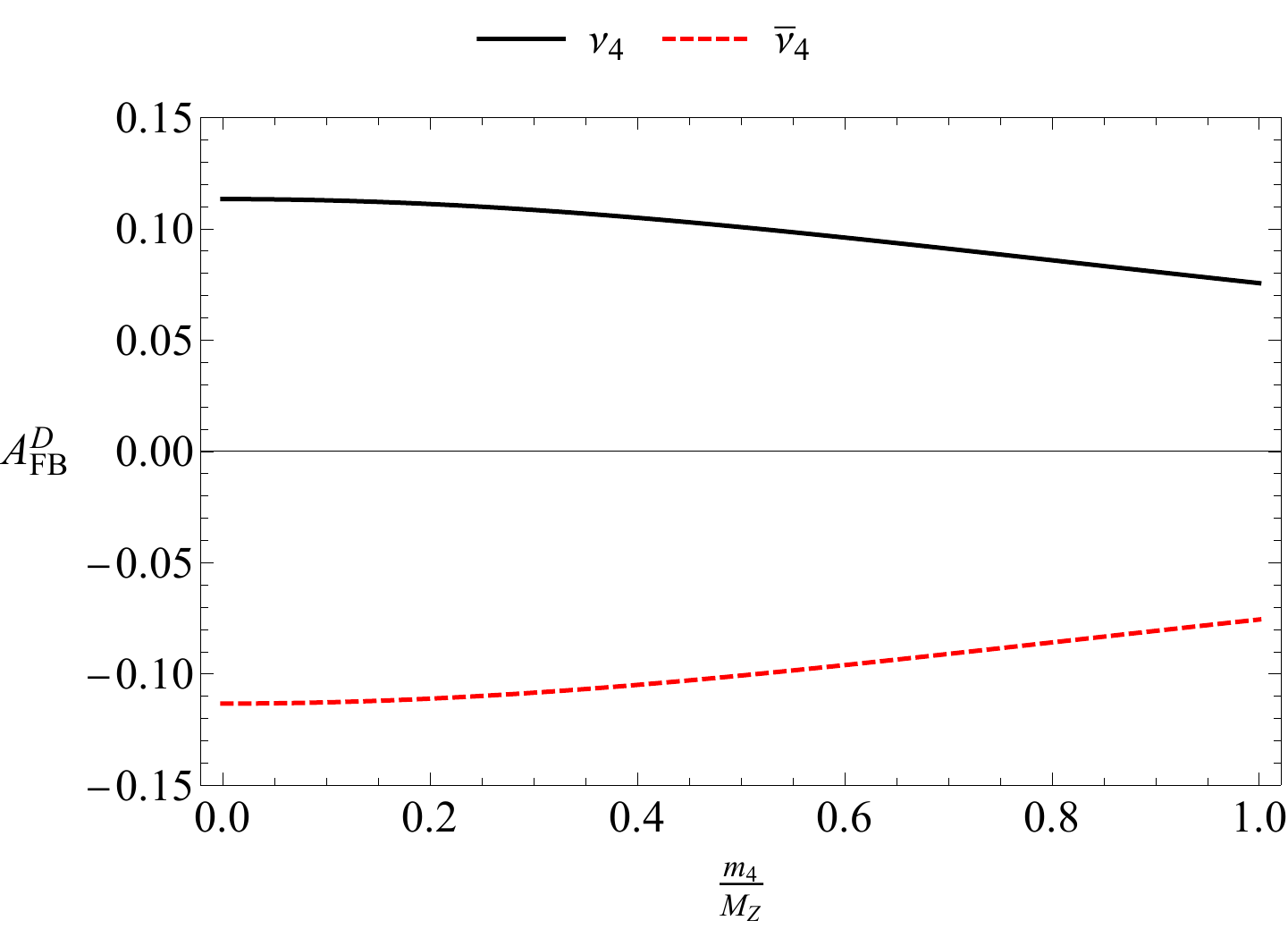}
\caption{\label{2a}}
\end{subfigure}\\
\begin{subfigure}{0.45\textwidth}
\includegraphics[scale=0.5]{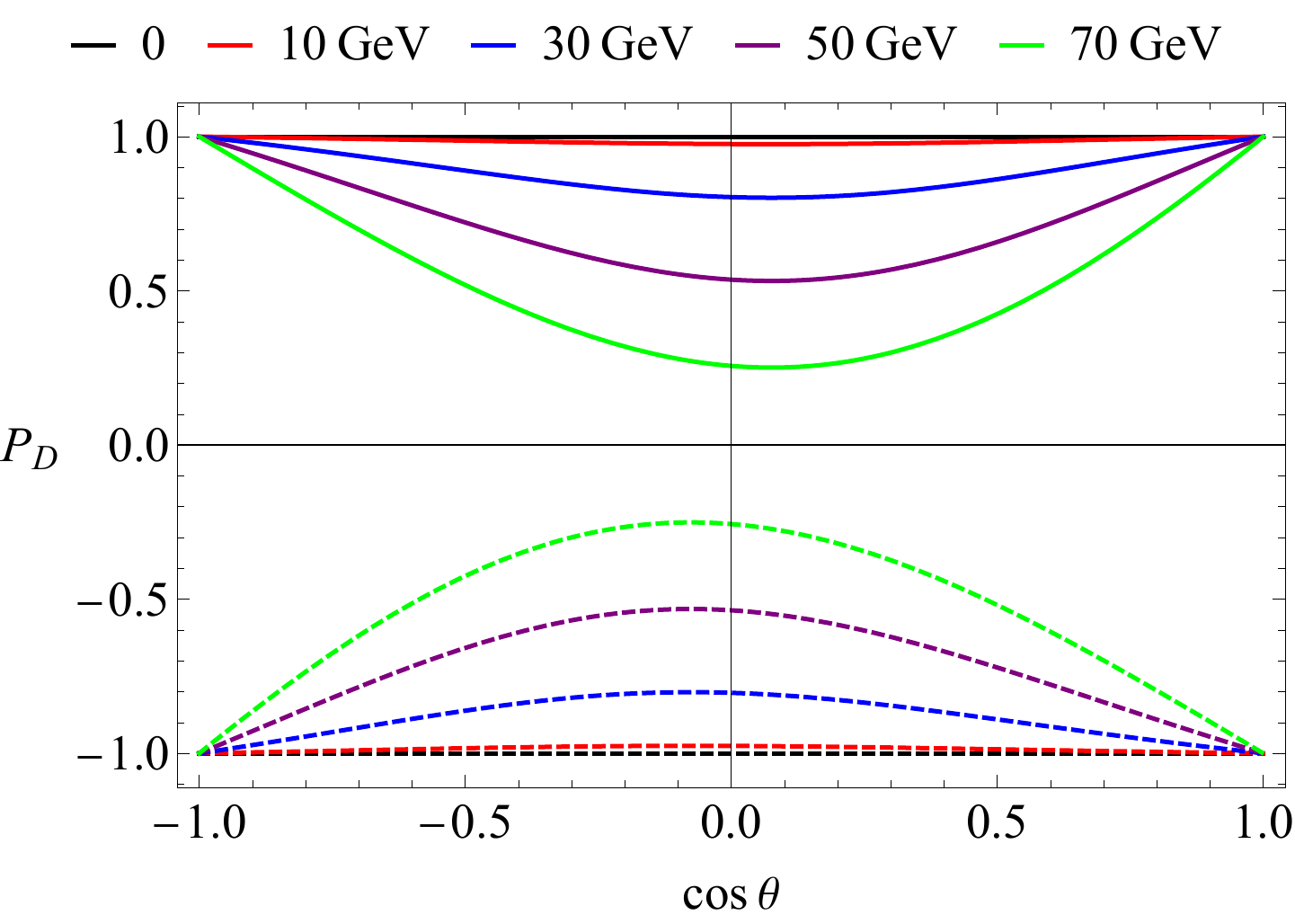}
\caption{\label{2b}}
\end{subfigure}
\begin{subfigure}{0.45\textwidth}
\includegraphics[scale=0.5]{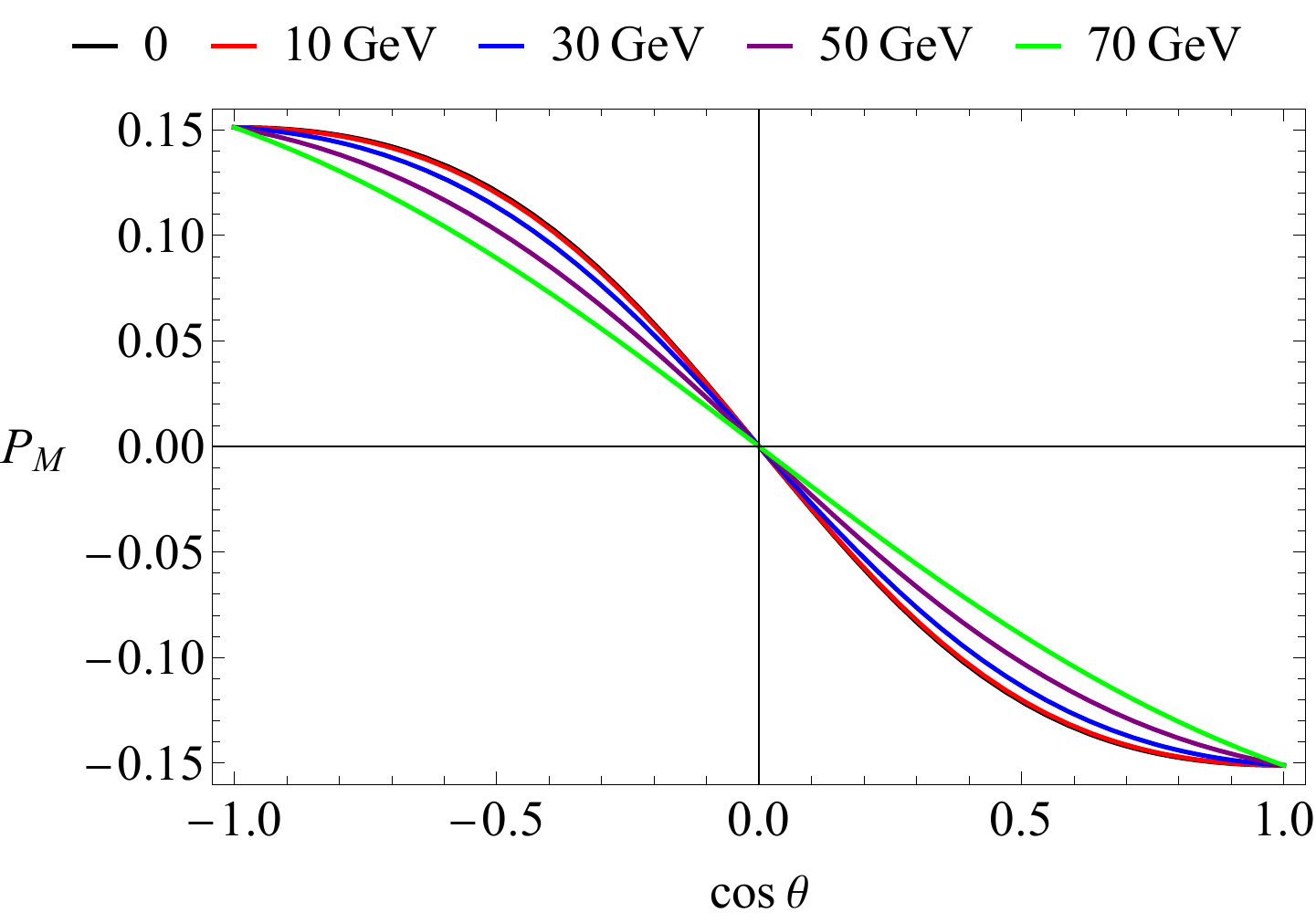}
\caption{\label{2c}}
\end{subfigure}
\end{center}
\caption{
\emph{(a)}: Forward-backward asymmetry as a function of $M/m_Z$.
\emph{(b) and (c)}:
Polarisations of Dirac ($P_D$) and 
 Majorana ($P_M$) HNLs as a function of the electron-HNL angle. 
 Plots from \cite{Blondel:2021mss}.
} \label{ForwardBackward}
\end{figure}

Collider studies are often performed in a \emph{phenomenological type I seesaw model}, defined by \eqref{eq:weak WW} with only one HNL species ($n=1$) of mass $M$.
This is not a realistic model of neutrino mass, but it can effectively capture many phenomenological aspects with only five parameters $(M,\theta_e,\theta_\mu,\theta_\tau,\Rll)$,\footnote{Practically it is often more convenient to consider the parameters $M$, $U^2=\sum_\alpha U_\alpha^2$ and the three ratios $\ratio{\alpha}=U_\alpha^2/U^2$, with $\alpha=e,\mu,\tau$. This also gives five parameters as $\sum_\alpha \ratio{\alpha} =1$. Note  that the $\theta_{\alpha i}$ for $n>1$ are in principle complex while the $U_{\alpha i}^2$ and  $\ratio{\alpha i}$ are real (and hence contain less information). However, the phases only play a role when there are interferences between the contributions from different $N_i$, which only occurs for $\Delta M \equiv |M_i-M_j| \sim \Gamma_N$, cf.~footnote \ref{JulietteQuestion}.} 
where $\Rll=0$ for Dirac-$N$ and $\Rll=1$ for Majorana-$N$. 
If all HNLs decay inside the detector the total number of events with $n=1$ is the same for the Dirac and Majorana cases,\footnote{\label{NumberOf EventsFootnote}Naively one may expect that the number of produced particles is twice as large for Dirac HNLs (compared to Majorana HNLs), reflecting the fact that Dirac fermions have twice as many internal degrees of freedom.
However, only half of them are produced in the decay of a given $Z$-boson (as $N$ is necessarily produced along with $\bar{\nu}$ and $\bar{N}$ along with $\nu$), 
and one can distinguish two possible types of final states that can be labeled by the light neutrino helicity.
The same is true for Majorana HNLs, hence $\cprod=1$ in both cases.
These conclusions are more general than the specific process considered here, cf.~e.g.~\cite{Drewes:2015iva,Bondarenko:2018ptm,Pascoli:2018heg,Ballett:2019bgd}.
The HNL decay rate $\Gamma_N$, on the other hand, is twice as large for Majorana HNLs with $\Rll=1$, as for Dirac HNLs ($\Rll=0$) the LNV processes are forbidden.
Hence, there are more possible final states for Majorana HNLs which are, however, indistinguishable when simply counting particles because the (anti)neutrino is not observed. Since all HNLs eventually decay, the total number of events is equal in both cases.} 
but there are at least three ways in which Dirac and Majorana HNLs can be distinguished at FCC-ee.
\begin{itemize}
    \item[1)]  In the Dirac case a $N$ ($\bar{N}$) is always produced along with a $\bar{\nu}$ ($\nu$). The 
    chiral nature 
    of the weak interaction and angular momentum conservation imply that $\nu$ and $\bar{\nu}$ are emitted with different angular distributions for a given $Z$-polarisation.
    Due to the parity-violation of the weak SM interaction the $Z$-bosons at lepton colliders are polarised at the level of 
    $P_Z\simeq 15\%$\footnote{$P_Z=(g_L^2-g_R^2)/(g_R^2 + g_L^2)\simeq 15\%$
    with $g_L=(1-2\sin^2\theta_W)$  and $g_R=2\sin^2\theta_W$ the left- and right-chiral neutral current charges of the charged leptons, respectively, and $\theta_W$ is the Weinberg angle \cite{Blondel:2021mss}.
    }
    even if the $e^\pm$ beams are not, 
    hence the angular distributions of the $N$ and $\bar{N}$ are different \cite{Blondel:2021mss}.\footnote{\label{AngularFootnote}
    For Dirac HNLs one finds differential production cross sections for $e^+e^-\to Z \to N \bar{\nu}$ and $e^+e^-\to Z \to \bar{N} \nu$ \cite{Blondel:2021mss}
    \begin{eqnarray}
\frac{1}{\sigma_{N,\bar{N}}}\frac{{\rm d}\sigma_{N,\bar{N}}}{{\rm d}c_\theta} = \frac{3}{4(g_R^2+g_L^2)}\frac{m_Z^2}{(2m_Z^2+M^2)}\left(g_R^2(1\mp c_\theta)^2+g_L^2(1\pm c_\theta)^2+ \frac{M^2}{m_Z^2}(g_R^2+g_L^2)s_\theta^2
\right)~, \nonumber 
\end{eqnarray}
with with $c_\theta$ and $s_\theta$ the sine and cosine of the angle between the HNL and electron momenta.
For Majorana HNLs  the angular distribution is  given by the sum of the differential $N$ and $\bar{N}$ production cross sections.}
    Since Dirac $N$ ($\bar{N}$) can only decay into leptons (antileptons), this introduces differences in the angular distribution of leptons and antileptons. This can be observed in the form of a forward-backward asymmetry $A_{FB}^D \simeq P_Z \frac{3}{4}/(1 - (M/m_Z)^2/2) \sim 10\%$, 
    cf.~Fig.~\ref{2a}. 
    For Majorana HNLs there is no forward-backward asymmetry because 
    they can decay into leptons and antileptons. 
\item[2)] For the Dirac case, the $N$ and $\bar{N}$ individually are highly polarised, cf.~Fig.~\ref{2b}, 
because $N$ ($\bar{N}$) can only have been produced along with $\bar{\nu}$ ($\nu$), whose helicity is fixed in the massless limit. Since $N$ can only decay into leptonic final states ($\bar{N}$ into antileptonic ones), the parent particle of leptons and antileptons tend to have opposite polarisation. The decay rates are polarisation-dependent \cite{Ballett:2019bgd}, leading to different spectra for leptons and antileptons \cite{Blondel:2021mss}.
For Majorana HNL there is no difference between $N$ and $\bar{N}$; their polarisaion is of order (and proportional to) $P_Z$, and they can decay into either leptons and antileptons.
This difference in the lepton spectra is observable.
\item[3)] For long-lived HNLs counting the number of events as a function of displacement provides an additional probe that is independent of $P_Z$. 
While the number $\NHNL$ of HNLs
produced in $Z$-decays 
along with a lepton or antilepton of flavour $\alpha$
is the same for Dirac and Majorana HNLs, 
their decay rate 
differs by a factor two,
leading to a twice larger decay length
in the detector $\lambda_{N} = \upbeta \upgamma/\Gamma_{N}$, with $\upbeta\upgamma= \pN/M$ and $\pN$ the HNL three-momentum.
Hence, the number of HNL decays into lepton flavour $\beta$ with a displacement between $l_0$ and $l_1$ is sensitive to this difference.
It is given by\footnote{
The simple analytic estimate \eqref{eq:observed events} can even describe the number of events in proton collisions surprisingly well if it is weighted by an appropriate momentum distribution that has to be obtained from simulations \cite{Bondarenko:2019yob,Drewes:2019vjy}. 
For the $Z$-pole run at lepton colliders 
\eqref{eq:observed events} is even more accurate, and the sensitivity region can be described analytically by \eqref{BlueLine}, \eqref{RedLine}, \eqref{GreenLine}, cf.~Fig.~\ref{1b}.
}
\begin{equation} \label{eq:observed events}
N_\text{obs} 
\simeq 
\ratio{\beta}
\NHNL
\left[\exp(-l_0/\lambda_{N}) - \exp(-l_1/\lambda_{N}) \right] \effa,
\end{equation}
with $0 \leq\effa\leq1$ an overall efficiency factor.\footnote{Neutral and charged current interactions allow for many possible final states $\X_\beta$ for given $\beta$, 
including leptonic decays
and semi-leptonic decays, cf.~\cite{Atre:2009rg}.
In practice one has to 
decompose $\effa$ into a sum of effective efficiency factors for each of them, $\effa = \sum_{\X_\beta} \epsilon_{\alpha\beta \X}$,
and factorise $\epsilon_{\alpha\beta \X}$
into the branching ratio $\BRb$ of the corresponding HNL decay and the 
actual detector efficiency factor $\epsilon_{\alpha\beta \X}^{\rm det}$ for that final state, $\epsilon_{\alpha\beta \X}=\BRb \epsilon_{\alpha\beta \X}^{\rm det}$.
If one is only interested in a specific sub-set of final states (as it is typically the case in a given experimental search), the effective efficiency factor can be utilised to select them by 
including only the desired $\X_\beta$ into the sum.
For the fully analytic treatment to be applicable, we neglect any dependence of 
$\effa$ on direction, energy or displacement, which can be justified as long as 
it leads only to errors of order one in $\Nobs$ (which affect the sensitivity 
region in figure \ref{1b} only mildly due to the steep dependence of 
\eqref{eq:observed events} on $M$ and $U^2$).}
Here\footnote{One advantage of using the ratios $\ratio{\alpha}$ is that experimental sensitivities can practically only be computed for fixed $\ratio{\alpha}$ and strongly depend on those ratios \cite{Drewes:2018gkc,Tastet:2021vwp}. Hence, benchmarks for experimental searches are typically defined by a choice of the $\ratio{\alpha}$ \cite{Beacham:2019nyx,Drewes:2022akb}. 
From a theoretical viewpoint using such ratios is convenient because the $\ratio{\alpha 1}+\ratio{\alpha 2}$ are in good approximation determined by light neutrino oscillation data alone for $n=2$ \cite{Hernandez:2016kel,Drewes:2016jae}, and in particular independent of $M$.} 
\begin{equation}
\ratio{\alpha}=\frac{U_\alpha^2}{U^2}\nonumber
\end{equation}
and \cite{Atre:2009rg,Antusch:2015mia}
\begin{equation}
\NHNL 
\simeq 
2
\ratio{\alpha}
U^2
\cprod
N_Z N_{\rm IP}
\BR
\Pi\nonumber
\end{equation}
where 
\begin{eqnarray}
\Pi=\left(\frac{2\pN}{m_Z}\right)^2
\left(
1+\frac{(M/m_Z)^2}{2}
\right) \ , \ \pN = \frac{m_Z}{2}\left(1 - (M/m_Z)^2\right), \nonumber
\end{eqnarray}
 $\BR=\operatorname{BR}(Z\mathpunct\to\nu_\alpha\overline\nu_\alpha) = \frac{1}{5} \frac{1}{3}$, 
 $\upbeta\upgamma= \pN/M$,
 $N_{\rm IP}$ and $N_Z$ the number of interaction points and number of $Z$-bosons produced at each of them,
and $\cprod$ 
 a numerical coefficient that is the same for Dirac or Majorana HNL ($\cprod=1$) if  $n=1$.
The decay rate is \cite{Atre:2009rg}
\begin{eqnarray}
\Gamma_{N} \simeq \cdec \frac{\aa}{96\pi^3}  U^2 M^5 G_F^2 \nonumber
\end{eqnarray}
with $\aa \simeq 12$
for $M < m_Z$ with $\cdec=1$ ($\cdec=1/2$) for Majorana (Dirac),\footnote{\label{JulietteActualQuestion}
For $n=1$ and assuming that the HNLs have no additional interactions, they either behave like Dirac particles ($\Rll=0$, $\cprod=1$, $\cdec=1/2$) 
or Majorana particles ($\Rll=1$, $\cprod=1$, $\cdec=1$). 
Hence, a determination of $\cdec$ with the lifetime method 3) in principle unambiguously answers the question which of these two options is realised in nature. 
In realistic models with $n>1$ the situation is more complicated, cf.~section \ref{sec:NeutrinoMassModels}.
In particular, as far as the lifetime method 3) is concerned, two Majorana HNLs with equal mixings $U_\alpha^2$ and a physical mass splitting $\Delta M$ that is smaller than the experimental resolution
$\delta M_{\rm exp}$  can appear like a Dirac HNL with an apparent mixing $2 U_\alpha^2$; they are effectively described by $\cprod=2$, $\cdec=1$ in the simple phenomenological model with $n=1$, cf.~footnote \ref{JulietteQuestion}.  
For $\Delta M \gg \Gamma_N$ 
these scenarios can be distinguished with methods 1) and 2), for $\Delta M \ll \Gamma_N$  
not.} 
and 
\begin{eqnarray}
\lambda_{N} = \frac{\upbeta \upgamma}{\Gamma_{N}}
\simeq \frac{1.6}{ U^2 \cdec} \left(\frac{M}{\rm GeV}\right)^{-6}\left(1-(M/m_Z)^2\right) {\rm cm}\nonumber.
\end{eqnarray}
\end{itemize}

\footnotetext{The small disagreement for $M < 5$ GeV can be fixed by replacing $\aa$ with a function that takes account of lepton and meson masses. Analytic approximations can e.g.~be found in~\cite{Bondarenko:2018ptm}.}

\section{Probing realistic neutrino mass models and leptogenesis}\label{sec:NeutrinoMassModels}
\begin{figure}
\centering
\begin{subfigure}{0.9\textwidth}
\centering
\includegraphics[scale=0.4]{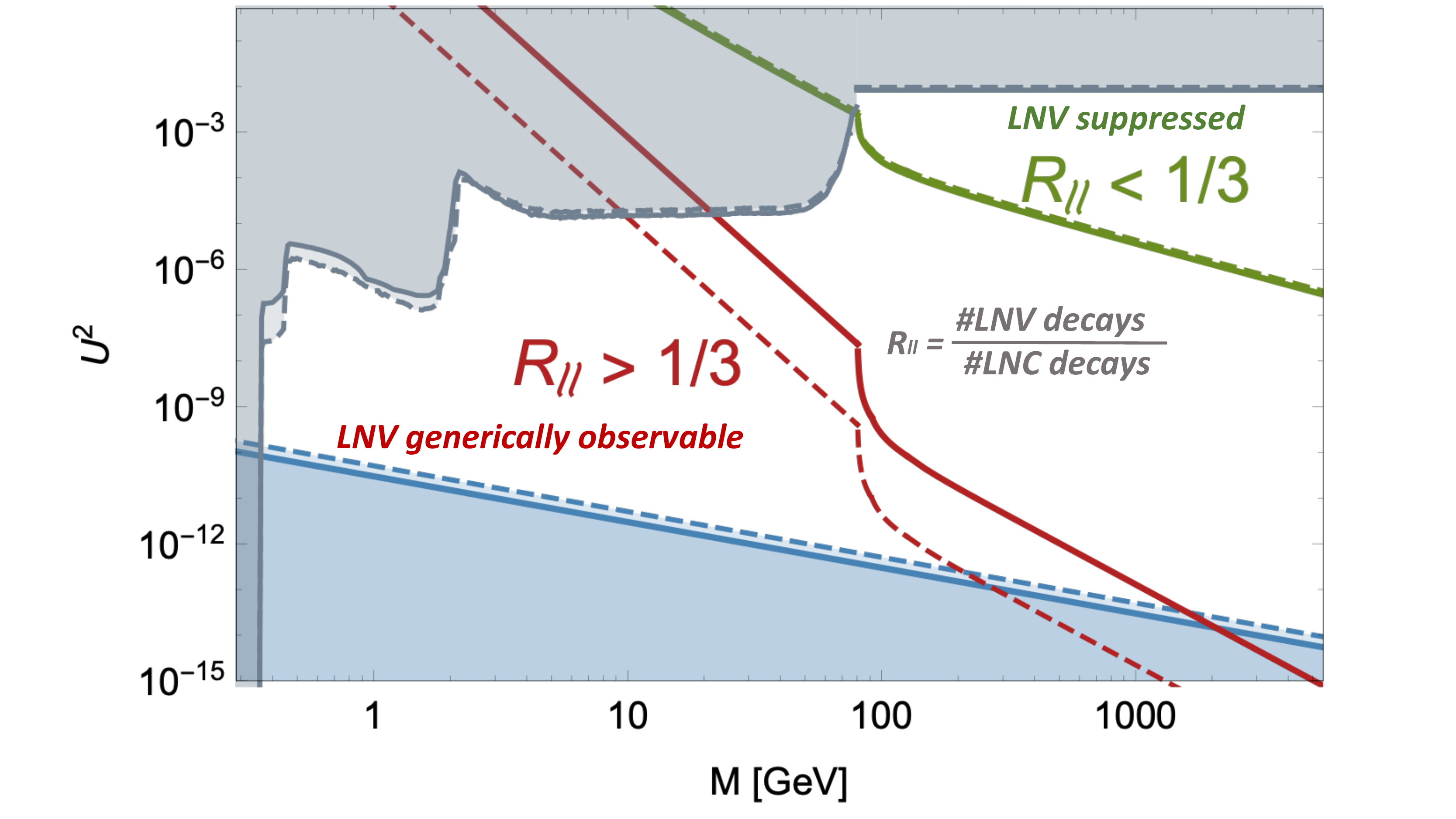}
\caption{\label{3a}}
\end{subfigure}\\
\begin{subfigure}{0.45\textwidth}
\includegraphics[scale=0.25]{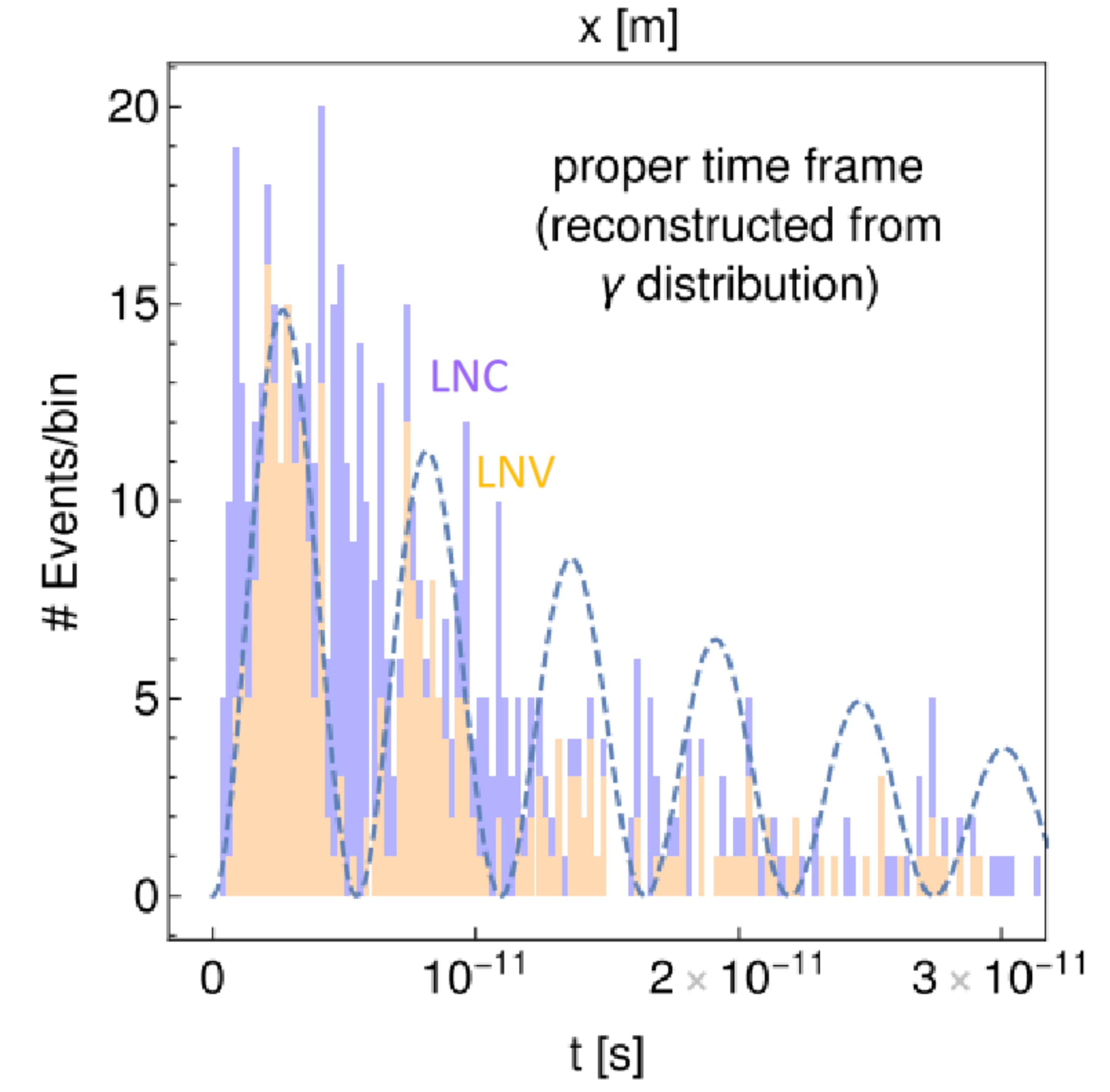}
\caption{ \label{3c}}
\end{subfigure}
\begin{subfigure}{0.45\textwidth}
\includegraphics[scale=0.25]{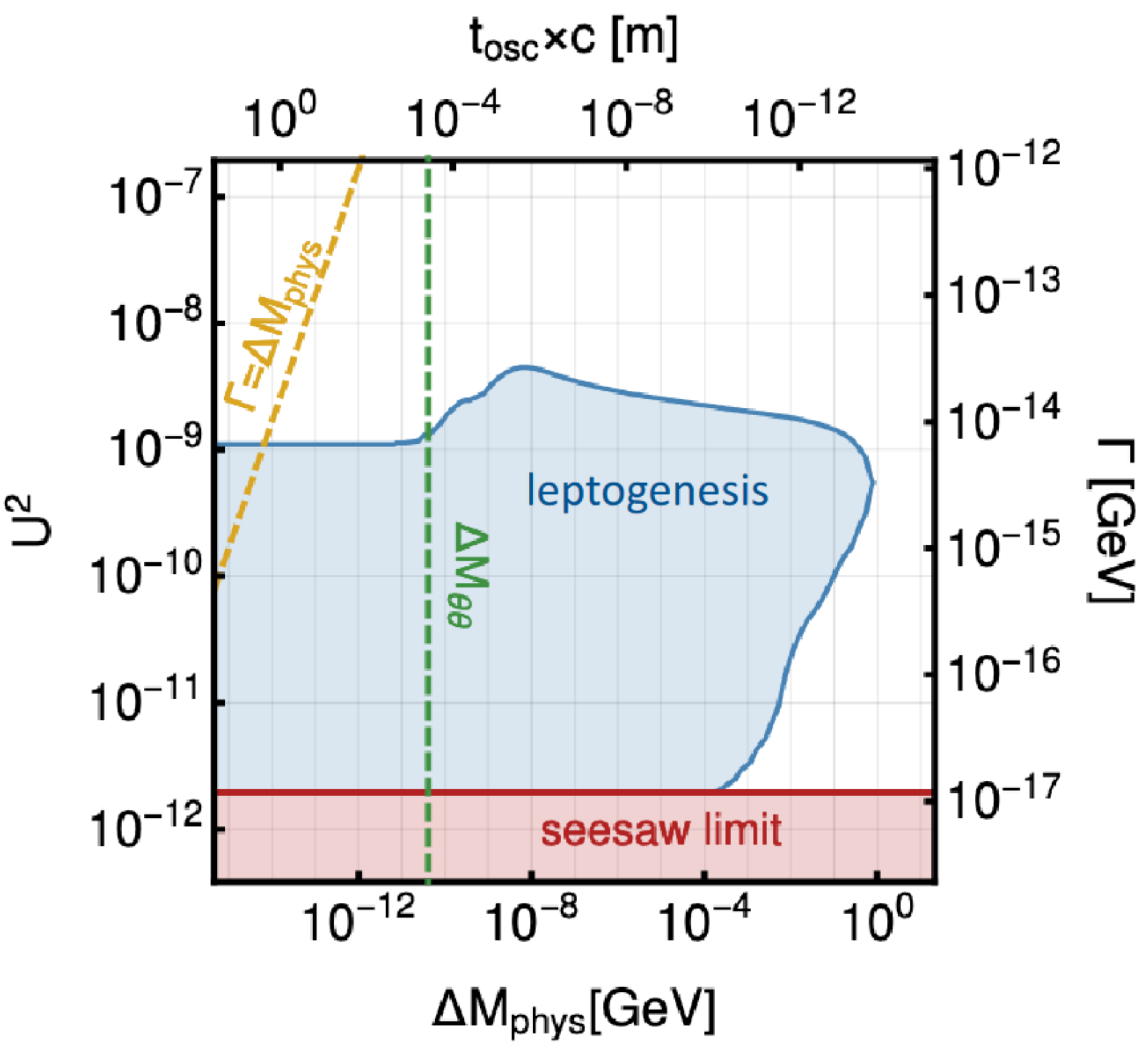}
\caption{ \label{3d}}
\end{subfigure}
\caption{\emph{(a)}:
Parameter region where $\Rll$ is suppressed or not \cite{Drewes:2019byd}, 
\emph{(b) and (c)}: reconstructed HNL oscillation time at LHCb \cite{Antusch:2017ebe} (will be easier at FCC-ee because of the smaller boost) compared to the oscillation time expected from leptogenesis \cite{Antusch:2017pkq}.
}
\label{LNVfigure}
\end{figure}

Realistic neutrino mass models typically require more than one HNL flavour.
In the type-I seesaw $n$ must equal or exceed the number of non-zero $m_i$. 
In  technically natural low-scale realisations  
that can be probed at colliders\footnote{A discussion of the motivation for low scale seesaw models can e.g.~be found in section 5.1 of \cite{Agrawal:2021dbo} and references therein.}
the $m_i$ are protected by a symmetry, cf~issue II).
If the symmetry is exact,  the HNLs  have to be organised in pairs 
with \cite{Moffat:2017feq}
\begin{displaymath}
M_i=M_j \ {\rm and} \  \theta_{\alpha i} = \ii \theta_{\alpha j}
\end{displaymath}
that form Dirac spinors  with distinctively different $N$ and $\bar{N}$; this would imply $m_i=0$ and $\Rll=0$.\footnote{\label{InterferenceFootnote}
Practically the symmetry manifests itself through destructive interference between the contributions from $N_i$ and $N_j$ to LNV processes \cite{Kersten:2007vk}.} 
Naively one would expect that the tiny symmetry breaking due to the $m_i\neq 0$ can only lead to an unobservably small $\Rll$, cf.~issue II).
However, even a small 
splitting $\Delta M$ between the physical HNL masses
induced by the symmetry breaking 
can give rise to $\bar{L}$-violating oscillations between the $N$-like and $\bar{N}$-like states inside the detector (cf.~\cite{Antusch:2020pnn,Antusch:2022ceb} and references therein). If the HNL decay length 
$\lambda_N$
exceeds the oscillation length, LNV processes are unsuppressed. 
Since $\Delta M \ll M_i$ for the approximate symmetry to protect the $m_i$, 
$\Delta M$ may be smaller than the experimental mass resolution $\delta M_{\rm exp}$,
resulting in a single resonance that is effectively characterised by a non-integer \cite{Anamiati:2016uxp}
\begin{equation}
\Rll = \frac{\Delta M^2 }{2\Gamma_N^2 + \Delta M^2}.\nonumber
\end{equation}
Figure \ref{3a} shows what values of $\Rll$ one can expect as a function of $M$ and $U^2$ \cite{Drewes:2019byd}, indicating that LNV would be observable in long-lived HNL searches at lepton colliders.
For $\Delta M/\Gamma_N\sim \ {\rm a \ few}$, it may further be possible to resolve the HNL oscillations by observing $\Rll$ as a function of the displacement \cite{Antusch:2017ebe}, cf.~Fig~\ref{3c}.

In summary, the phenomenology of realistic seesaw models is much richer than that of the widely-used phenomenological model \eqref{eq:weak WW} with $n=1$. Effectively many aspects can effectively still be captured in this model by adjusting $\Rll$, 
$\cdec$, 
$\cprod$ to non-integer values,\footnote{\label{cDefs}
We define $\cprod$ and $\cdec$ in a way that the 
mixings in \eqref{eq:observed events} refer to those of one state, i.e., one should replace $U_\alpha^2\to U_{\alpha i}^2$ and $U^2\to U_i^2$ in $\Gamma_N$, $\NHNL$, and the definitions of the ratios $\ratio{\alpha}$ (then $\ratio{\alpha i}$). } 
and if one considers $\Rll$ as a function of the displacement.
The extreme cases $\Rll=1$ and $\Rll=0$ are realised in the lower left and upper right corner of Fig.~\ref{3a}, respectively.
They represent well-defined benchmarks \cite{Drewes:2022akb} that can easily be implemented in event generators,\footnote{Some of the most widely used tools \cite{Alva:2014gxa,Degrande:2016aje,Coloma:2020lgy} have implemented these benchmarks. Simulating HNL oscillations inside the detector is not foreseen in existing event generators; a 
$\mathtt{FeynRules}$ \cite{Alloul:2013bka} model file 
and a patch for $\mathtt{MadGraph}$ \cite{Alwall:2011uj}
that permit to effectively treat them in  have recently been developed in \cite{Antusch:2022ceb}.}
but it is important to keep in mind that nature is likely to be more complex.\footnote{\label{JulietteQuestion}
Limiting cases for $n=2$ can effectively be described by \eqref{eq:observed events}, \eqref{RedLine},  \eqref{GreenLine}, \eqref{BlueLine} with 
\begin{eqnarray}\nonumber
\begin{tabular}{c c c c c}
mass spectrum & $\cprod$ & $\cdec$ & $\Rll$ & appearance \\
 \hline
$\Delta M > \delta M_{\rm exp} \gg \Gamma_N$ & $1$ & $1$ & $1$ & two Majorana HNLs with mixing $U^2$ each\\
$  \delta M_{\rm exp} > \Delta M \gg \Gamma_N$ & $2$ & $1$ & $1$ & one HNL, mixing $2U^2$, lifetime as Dirac, $\Rll$ as Majorana\\
$  \delta M_{\rm exp} > \Gamma_N \gg \Delta M$ & $2$ & $1$ & $0$ & one Dirac HNL with mixing $2U^2$\\
\end{tabular}
\end{eqnarray}
Here $\delta M_{\rm exp}$ is the experimental mass resolution and we follow the conventions from footnote \ref{cDefs}.
Note that there is no suppression of the overall number of events for $ \Delta M / \Gamma_N \to 0$ in spite of $\Rll \to 0$ because the destructive interference in the $L$-violating channel is accompanied by a constructive interference in the $L$-conserving channel (and also no change in the decay length $\lambda_N$, in contrast to the case with $n=1$ sketched in footnote \ref{NumberOf EventsFootnote}).
Here we assume that the phases are chosen such that $\theta_{\alpha i}={\rm i} \theta_{\alpha j}$, as dictated by the $U(1)_{B-\bar{L}}$ symmetry; deviations from this or intermediate values of $\Delta M/\Gamma_N$ lead to modifications \cite{Abada:2022wvh}.
 If more than two HNLs have quasi-degenerate masses 
the situation is even more complicated \cite{Drewes:2019byd}.
}

\section{Practical feasibility and number of events}
Discovering HNLs only requires a handful of events, but studying their properties with the methods 1)-3) (or others) will require reliable statistics. In displaced vertex searches the number of events can vary by many orders of magnitude across the sensitivity region. For the $Z$-pole run at FCC-ee or CEPC we can reliably estimate the total number of observed HNL decays 
inside a cylindrical detector of length $l_{\rm cyl}$ and diameter $d_{\rm cyl}$
by 
identifying $l_0$ in \eqref{eq:observed events} with the smallest displacement for which the assumption of vanishing backgrounds can be justified\footnote{
Of course HNLs can also be searched for in prompt decays, and mixings $U^2 > U^2_{\rm max}$ may be accessed experimentally \cite{BayNielsen:2017yws,Shen:2022ffi}, 
but our simple approach cannot be applied because SM backgrounds need to be taken into account, cf.~e.g.~\cite{Alimena:2019zri,MATHUSLA:2020uve,Knapen:2022afb} and references therein for recent discussions on the assumption of background-freedom.} 
and
setting
\begin{equation}
 l_1 = \frac{1}{2} (3/2)^{1/3} 
 d_{\rm cyl}^{2/3} l_{\rm cyl}^{1/3}\nonumber
 \end{equation}
 so that a sphere of radius $l_1$ has the same volume as the cylinder, cf.~Fig.~\ref{fig:illustrative plot}. 
 In the limit of an infinitely large detector we can estimate the
maximal mixing $U^2_{\rm max}$  
and the minimal mixing $U^2_{\rm min}$ 
for which one can see $\Nobs$ events
by solving \eqref{eq:observed events} with $l_1\to\infty$ for $U^2$,
\begin{subequations}
 \begin{align}
    U^2_{\rm min} &= \frac{W_{0}\left(
    X Y
    \right)}{X} 
     \simeq Y
     \label{RedLine}\\ 
      U^2_{\rm max} &= \frac{W_{-1}\left(
    X Y
    \right)}{X}
    \simeq 
      \frac{\log\left(-X Y
    \right)}{X}
    \label{GreenLine}
\end{align}
\end{subequations}
where 
\begin{equation}
X =
-\frac{l_0}{U_\beta^2 \lambda_N}
=
-\frac{\aa  G_F^2 l_0 M^6  \cdec }{96\pN \pi^3} \ , \quad
Y=
\frac{U^2 \Nobs/\ratio{\beta}}{\effa \NHNL}
= \frac{N_\text{obs}/(\ratio{\alpha}\ratio{\beta})}{2 \effa \BR \cprod N_{\rm IP} \Pi N_Z},\nonumber
\end{equation}
with $W_s$ the $s$-branch of the Lambert W-function. 
For $U^2 > U^2_{\rm max}$ for 
assumption of background-freedom is not justified. For $U^2 < U^2_{\rm min}$ less than $\Nobs$ HNLs are produced in the first place, so even an infinitely large ideal detector could not see enough decays. Both limits strongly depend on $M$.
The finite detector size comes into play for very long-lived HNLs, for which one can expand the exponential in \eqref{eq:observed events} and find (neglecting $l_0$)
\begin{equation}\label{BlueLine}
     U_{\rm min}^2 
     = \frac{2^{1/6} 3^{1/3} 8 \pi^{3/2}(\pN Y)^{1/2}}{
     (\aa  \cdec )^{1/2}
     G_F M^3
     d_{\rm cyl}^{1/3} l_{\rm cyl}^{1/6}
     }
     \simeq 
     \sqrt{\frac{N_\text{obs}}{\ratio{\alpha}\ratio{\beta}}} 
     \frac{57}{G_F M^3}\sqrt{\pN}
     d_{\rm cyl}^{-1/3} l_{\rm cyl}^{-1/6}
     \left(
     \effa N_Z
     N_{\rm IP}
     \cdec 
     \cprod
     \Pi
     \right)^{-1/2}
\end{equation}
The dependence of \eqref{BlueLine} on $l_{\rm cyl}$ and $d_{\rm cyl}$ quantifies the sensitivity gain with additional detectors \cite{Chrzaszcz:2020emg,Wang:2019xvx}.
The smallest mixing that can be probed is given by the maximum of $U_{\rm min}^2$ 
in \eqref{RedLine} 
and \eqref{BlueLine}; one can estimate the point where they cross at 
   \begin{equation}
M \simeq 
\frac{2.75}{G_F^{1/3} 
    d_{\rm cyl}^{1/9} l_{\rm cyl}^{1/18}
    } 
    \left(
    \frac{\BR \cprod N_Z m_Z N_{\rm IP}}{\cdec  N_\text{obs}/(\ratio{\alpha}\ratio{\beta})}
    \right)^{1/6}
\simeq
\lim_{M/m_Z\to 0} 2.75 \left(\frac{\pN}{Y G_F^2 \cdec d_{\rm cyl}^{2/3} l_{\rm cyl}^{1/3}}
\right)^{1/6}
    .\nonumber
    \end{equation}
    Since $\NHNL \propto U^2$
    one can potentially see over a million events at FCC-ee or CEPC, cf.~Fig.~\ref{fig:illustrative plot}.\footnote{Fig.~\ref{fig:illustrative plot} shows that \eqref{eq:observed events}, \eqref{BlueLine}, \eqref{RedLine}, \eqref{GreenLine} are very accurate for Majorana HNLs. One reason for this is that the HNLs are in good approximation emitted isotropically in the $Z$-boson rest frame, and the $Z$s practically decay at rest in the laboratory frame. For Dirac HNLs there are larger deviations from isotropy (cf.~Fig.~\ref{ForwardBackward} and footnote \ref{AngularFootnote}), but one can still expect the analytic relations \eqref{eq:observed events}, \eqref{BlueLine}, \eqref{RedLine}, \eqref{GreenLine}  to provide good approximations because changes in $N_\text{obs}$ by factors of order one only mildly affect the sensitivity region in figure \ref{1b} due to the steep dependence of \eqref{eq:observed events} on $M$ and $U^2$.
    This observation may be generalised to other types of long-lived particles. For instance, the axion-like particles (ALPs) discussed in section 2.2 of \cite{Blondel:2022qqo}  (cf.~\cite{Bauer:2018uxu}) can also be produced in $1\to 2$ decays of $Z$-bosons, and the angular dependence of the production cross section is only of order one ($\sim 1 + c_\theta^2$).
    Hence one should be able to derive analogous relations to those presented here by replacing $2\ratio{\alpha}U^2\cprod\BR$ with the corresponding branching ratio of $Z$-decays into ALPs, $M_N$ with the ALP mass, $\Gamma_N$ with the total ALP decay width, and $\ratio{\beta}$ with the ALP decay branching ratio into the final state under consideration.
    }
This does not only make the methods 1)-3) to search for LNV feasible,\footnote{For instance, one would need $\sim 10^2$ events to rule out a $\sim 10\%$ forward backward asymmetry, cf.~Fig.~\ref{2a}.
Fig.~\ref{1b} shows that this is possible with mixings as small as $U^2\sim 10^{-9}$.} 
but also allows for further measurements of the HNL properties, including measurements of $\Rll$ and of the $\ratio{\alpha}$.
The sensitivity gain that can be achieved with additional detectors \cite{Chrzaszcz:2020emg} can be estimated with \eqref{BlueLine}. 
This shows the potential of lepton collider to test neutrino mass models and leptogenesis \cite{Antusch:2017pkq}.

\section*{Acknowledgements}

I would like to thank all members of the informal \emph{Long-lived particles at the FCC-ee} working group for discussions that lead to this contribution, in particular Juliette Alimena, Alain Blondel, Rebeca Gonzalez-Suarez, and Suchita Kulkarni. I also thank Yannis Georis, Jan Hajer, Juraj Klaric and Maksym Ovchynnikov for helpful discussions.

\bibliographystyle{JHEP}
\bibliography{bibliography}{}
\end{document}